\documentclass[aps,twocolumn,pre]{revtex4-2}
\usepackage{amsthm}
\usepackage{ulem}
\usepackage{graphicx}
\usepackage{amsfonts}
\usepackage{amsmath}
\usepackage{bbold}
\usepackage{bm}
\usepackage{enumerate}
\usepackage{color}
\begin{document}

\title{Trade-off between coherence and dissipation for excitable phase oscillators}
\author{Chunming Zheng}
\affiliation{School of Physics and Astronomy, Yunnan University, Kunming, 650091, China}

\begin{abstract}
Thermodynamic uncertainty relation (TUR) bounds coherence in stochastic oscillatory systems. In this paper, we show that both dynamical and thermodynamic bounds play important roles for the excitable oscillators, e.g. neurons. Firstly, we investigate the trade-off between coherence and dissipation both in the sub and super-threshold regions for a single excitable unit, where both the TUR and the SNIC bounds constrain the fluctuation of inter-spike intervals. Secondly, we show that the widely studied phenomenon called coherence resonance, where there exists a noise strength to make the oscillatory responses of the system most coherent, is also bounded by the TUR in the one-dimensional excitable phase model. Finally, we study the coherence-dissipation relation in ensembles of strongly coupled excitable oscillators.
\end{abstract}
\maketitle
\section{Introduction}
Thermodynamic constraints on the fluctuations of currents in non-equilibrium steady states have attracted lots of attention in recent years. The thermodynamics uncertainty relation \cite{barato2015thermodynamic}, first discovered in the continuous Markov processes, states that there is an trade-off inequality relation between the precision of non-equilibrium currents and the entropy production of the system. It implies that one has to pay more thermodynamic cost to achieve higher accuracy of fluctuating currents.
Since its discovery, the TUR has been investigated and applied in a variety of settings \cite{barato2016cost,gingrich2016dissipation,pietzonka2016universal,garrahan2017simple,gingrich2017fundamental,neri2017statistics,macieszczak2018unified,lee2018thermodynamic,hasegawa2019fluctuation,hasegawa2020quantum,liu2020thermodynamic,falasco2020unifying,pearson2021measuring,dechant2021continuous,hasegawa2021thermodynamic,yoshimura2021thermodynamic,pietzonka2022classical,van2023thermodynamic,dieball2023direct,pietzonka2024thermodynamic,dechant2018current}, among which are both classical and quantum systems, time-independent and time-dependent systems and so on.

The first-passage-time (FPT) fluctuation, serving as the conjugate problem of current fluctuation, also follows the TUR \cite{gingrich2017fundamental,neri2017statistics,garrahan2017simple}. For biochemical oscillators, it was reported that more entropy production is needed to achieve higher accuracy of oscillation \cite{cao2015free,barato2016cost,fei2018design}. While many works have been done for the  biochemical oscillators, it remains poorly understood for the excitable oscillators such as neurons, where there may exist fixed points in the neuronal dynamics without noisy excitation. If there is no external perturbation, the system remains in the rest state, while it's in the excited or firing state when it's perturbed by external force such as noise. After excitation, the system is usually accompanied by a refractory period before coming back to the rest state. Another firing event occurs after a large excursion of the system’s variables, e.g. the membrane voltage of the neurons, through phase space. 
Bifurcation usually occurs from a fixed point representing the rest state, to coherent oscillatory spiking in neurons \cite{izhikevich2007dynamical}. Furthermore, it has been demonstrated that the coherence in the neuronal systems has constraints by itself. For instance, the fluctuation of the inter-spike-interval (ISI) does not fall below the number $1/3$ for the quadratic integrate-and-fire neuron \cite{lindner2003analytic}, or equally the Theta neuron \cite{ermentrout1986parabolic}, and the active rotator \cite{sigeti1989pseudo,lindner2004effects}, where a nearly standard SNIC (saddle-node on an invariant circle) bifurcation occurs. 

Another typical example of constraint of coherence in excitable systems is manifested by coherence resonance \cite{pikovsky1997coherence}, where an optimal noise strength maximizes the coherence of oscillations such as the neuronal oscillations. Coherence resonance has been studied not only theoretically but also experimentally in various systems \cite{kiss2003experiments,ushakov2005coherence,lee2010coherence,mompo2018coherence}. 
Considering the TUR mentioned above, a natural question arises that whether these two constraints of coherence and the constraint due to thermodynamics, i.e. the TUR, can be investigated in the same setting.

To this end, we consider throughout this paper a generic phase oscillator model, which can be considered as a minimum one for the excitable system. Noteworthy, the active rotator model to be considered in the following, has been widely used to describe qualitatively complex phenomena in biological realistic neuronal models, e.g. the Hodgkin-Huxley type model \cite{kromer2016emergence,zheng2018delay}, and to mimic human cortex dynamics \cite{buendia2021hybrid}.
\section{Coherence-dissipation bound for a single excitable unit}
\label{Sec:single_oscillator}
We consider an excitable system that can be described by its phase dynamics
\begin{equation}
    \dot{\theta}=a+f(\theta)+\xi(t),
    \label{Eq:Model}
\end{equation}
where $a$ is a constant force and can represent the excitability of the system. Here $f(\theta)=-V'(\theta)$ is a periodic function with $f(0)=f(2\pi)$.
$\xi(t)$ is the Gaussian white noise with mean $\langle\xi(t)\rangle=0$ and variance $\langle\xi(t)\xi(t')\rangle=2D\delta(t-t')$. The system is excitable when $a<|f(\theta)|$ and is in the oscillatory state for $a>|f(\theta)|$.

A spike event can be defined as the phase crosses a given value, e.g. $\pi/2$. To quantify the non-equilibrium thermodynamics in the model \eqref{Eq:Model}, we write down the entropy production rate of generating one spike, following \cite{seifert2012stochastic},
\begin{equation}
    \dot{S}=\int_0^{2\pi}\frac{J^2}{DP(\theta)}d\theta.
\end{equation}
Here the probability current $J$ in the stationary state is a constant for one-dimensional noisy systems \cite{risken1996fokker}.
The mean inter-spike interval (ISI), i.e. equally the mean first-passage time (MFPT) from $0$ to $2\pi$, has the following expression
\begin{equation}
    \langle T\rangle=\frac{2\pi}{\langle\dot{\theta}\rangle}=\frac{2\pi}{\langle a+f(\theta)\rangle}.
    \label{Eq:MeanISI}
\end{equation}
As the probability $P(\theta)$ obeys the Fokker-Planck equation
\begin{equation}
\begin{aligned}
    \frac{\partial P(\theta)}{\partial t} =& -\frac{\partial}{\partial\theta}J\\
    =&-\frac{\partial}{\partial\theta}[a+f(\theta)]P(\theta) + D\frac{\partial^2}{\partial\theta^2}P(\theta),
\end{aligned}
\end{equation}
we have
\begin{equation}
    (a+f(\theta))P_{st}(\theta)=J+D\frac{\partial}{\partial\theta}P_{st}(\theta).
    \label{Eq:Flux}
\end{equation}
Here $P_{st}(\theta)$ is the stationary phase distribution.
Plugging \eqref{Eq:Flux} into \eqref{Eq:MeanISI} and because of the periodic boundary condition we arrive at
\begin{equation}
    \langle T\rangle=\frac{1}{J}.
\end{equation}
The entropy production during the MFPT $\langle T\rangle=1/J$ is therefore given by
\begin{equation}
    \sigma=\dot{S}\langle T\rangle=\int_0^{2\pi}\frac{J}{DP(\theta)}d\theta.
    \label{Eq:EntropyProd1}
\end{equation}
Then, substituting the expression of $J$, i.e. \eqref{Eq:Flux}, into the above equation \eqref{Eq:EntropyProd1} and due to the periodic boundary condition we obtain
\begin{equation}
    \sigma=\int_0^{2\pi}\frac{(a+f(\theta))P_{st}(\theta)-DP'_{st}(\theta)}{DP_{st}(\theta)}d\theta=\frac{2a\pi}{D}.
    \label{Eq:EntropyProd}
\end{equation}
This simple final expression implies that the entropy production during the time $\langle T\rangle$ has nothing to do with whether or not there are fixed points on the circle and is only related to the excitability $a$ and the noise intensity $D$.
The mean and variance of the ISI, i.e. the
first-passage time $T$ for the Langevin Eq.~\eqref{Eq:Model}, are given by quadrature expressions \cite{reimann2001giant} in the stationary state
\begin{equation}
    \langle T\rangle=\frac{\int_0^{2\pi}d\theta I_{\pm}(\theta)}{1-e^{-2\pi a/D}}, \quad 
    \langle \Delta T^2\rangle=\frac{2D\int_0^{2\pi}d\theta I_{+}(\theta)I_{-}^2(\theta)}{(1-e^{-2\pi a/D})^3},
    \label{Eq:Mean_var_ISI}
\end{equation}
where the function $I_{\pm}$ represents
\begin{equation}
    I_{\pm}=\frac{1}{D}e^{\pm V(\theta)/D}\int_{\theta-2\pi}^{\theta}dy e^{\mp V(y)/D}.
    \label{Eq:I_pm}
\end{equation}
Note that these quadrature formulas have to be evaluated numerically to be precise.
The fluctuation of the ISI is given by
\begin{equation}
\begin{aligned}
    F&=\frac{\langle\Delta T^2\rangle}{\langle T\rangle^2}\\
    &=\frac{D}{\pi(1-\exp(\frac{-2\pi a}{D}))}\frac{\langle I_{+}^2(\theta)I_{-}(\theta)\rangle}{\langle I_{+}(\theta)\rangle^2}.
\end{aligned}
\label{Eq:Fluc_ISI}
\end{equation}
The second equality is obtained by direct substitution of \eqref{Eq:Mean_var_ISI}.
Plugging the expressions of $\langle T\rangle$, $\langle\Delta T^2\rangle$ and replace $\frac{2\pi a}{D}$ by the entropy production $\sigma$, one can obtain the fluctuation of the ISI as the function of the entropy production $\sigma$ and evaluate it numerically. 

\begin{figure}
\centering
\includegraphics{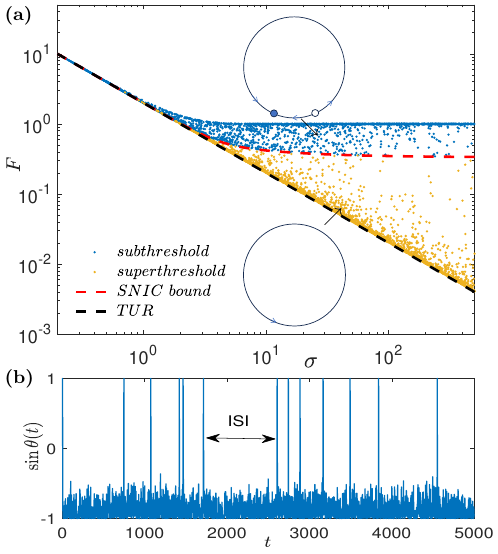}
\caption{(a) Coherence vs the entropy production with random parameters $a$ and $D$ in the subthreshold region (blue dots) and superthreshold region (yellow dots). The insets are schematics of the two regions in the deterministic case, where there exist one stable and unstable fixed points represented by the filled and empty circle respectively in the subthreshold region, and no fixed point in the superthreshold region. (b) A typical spike train generated by the model \eqref{Eq:Model} in the excitable region with $f(\theta)=\sin\theta, a=0.9$ and $D=0.02$. An inter-spike interval (ISI) is defined as the time needed to go once around the circle.}
\label{Fig:TUR}
\end{figure}

In this section we consider the canonical form of type-I excitable systems, i.e. the so called active rotator model \cite{shinomoto1986phase} or the Adler equation \cite{adler1946study} with $f(\theta)=\sin\theta$, which is a simplest model to investigate the type-I excitable systems. Many complex neuronal phenomena can be investigated qualitatively by this simplified model. In the deterministic case, there exist one stable fixed point (node) $\theta=\arcsin(a)-\pi$ and one unstable fixed point (saddle) $\theta=-\arcsin(a)$ for $a<1$. As $a$ becomes larger and surpass the critical point $a=1$, the system undergoes a SNIC bifurcation, resulting in an oscillatory state. Close to the bifurcation point $a=1$, the system is to the second order approximation of the normal form of saddle-node bifurcation and is closely related to the Theta-neuron model \cite{ermentrout1986parabolic} or the quadratic integrate-and-fire neuron. The schematic in the deterministic case is shown in the inset of Fig.~\ref{Fig:TUR} (a).

Due to the excitation of noise, a spike event that can be defined as the periodic phase $\theta$ crosses a prescribed value, e.g. $\pi/2$, occurs. So the ISI statistics is associated with the first-passage time problem.
We plot the fluctuation of the ISI versus the the entropy production as shown in Fig.~\ref{Fig:TUR} (a), where we choose random noise intensity $D$ and random $a$ in excitable (subthreshold) region ($a<1$) and oscillatory (superthreshold) region ($a>1$), respectively. We see that the excitable and oscillatory region is separated by the $F-\sigma$ curve of the critical state $a=1$. We call this curve the SNIC bound because of the SNIC bifurcation occurring here in the deterministic case. The SNIC bound can serve generally as the lower bound for the excitable region and the upper bound for the oscillatory region.
Furthermore, on this bound $a=1$, $\langle I_{+}^2(\theta)I_{-}(\theta)\rangle/\langle I_{+}(\theta)\rangle^2\rightarrow\frac{\pi}{3D}$ \cite{reimann2002diffusion}, so the fluctuation of the ISI described by Eq.~\eqref{Eq:Fluc_ISI} approaches $1/3$ as the entropy production increases. This is consistent with previous results in systems of standard SNIC bifurcation \cite{sigeti1989pseudo,lindner2004effects}. It means the coherence in the excitable region is limited no matter how much entropy production the system produces. 
In the small entropy production limit, i.e. when the noise is large, $\langle I_{+}(\theta)\rangle\rightarrow2\pi/D$, $\langle I_{+}^2(\theta)I_{-}(\theta)\rangle\rightarrow8\pi^3/D^3$, and thus $F\rightarrow\frac{2}{\sigma}$, which is close to the bound defined by the TUR \eqref{Eq:TUR} to be described in the following. The fluctuation of the ISI at the criticality $a=1$ thus boils down to the following expression:
\begin{equation}
    F=\begin{cases}
          \frac{1}{3}, &\sigma\rightarrow\infty,\\\\
           \frac{2}{\sigma}, &\sigma\rightarrow 0.
       \end{cases}
\end{equation}

The thermodynamic uncertainty relation (TUR) of the ISI, i.e. the first-passage time for the phase crosses a $2\pi$ period, according to Ref.~\cite{gingrich2017fundamental} reads
\begin{equation}
    F\geq \frac{2}{\sigma},
    \label{Eq:TUR}
\end{equation}
where the entropy production $\sigma$ during the MFPT $\langle T\rangle$ is given by Eq.~\eqref{Eq:EntropyProd}. We consider the TUR as a proven result and detailed proof is referred to \cite{gingrich2017fundamental}. As shown in Fig.~\ref{Fig:TUR} (a), the TUR bound is tight in the superthreshold region, while in the subthreshold region the bound becomes loose as the entropy production increases to some extent. In this sense, for large $\sigma$ or low noise intensity, the SNIC bound offers better constraint bound in the subthreshold region than the TUR does. Note that the fluctuation $F$ will approach other value instead of $1/3$ for other choices of the function $f(\theta)$, but the general $F-\sigma$ diagram with the critical bound separating the subthreshold and superthreshold regions will be similar to that in Fig.~\ref{Fig:TUR} (a). A typical spike train generated by the model \eqref{Eq:Model} in the excitable (subthreshold) region is shown in Fig.~\ref{Fig:TUR} (b).

Another well-studied phenomenon in excitable systems is coherence resonance, which
we will revisit through the lens of thermodynamic constraint in the following section.

\section{Thermodynamic-bounded coherence resonance}
\begin{figure}[t]
\centering
\includegraphics{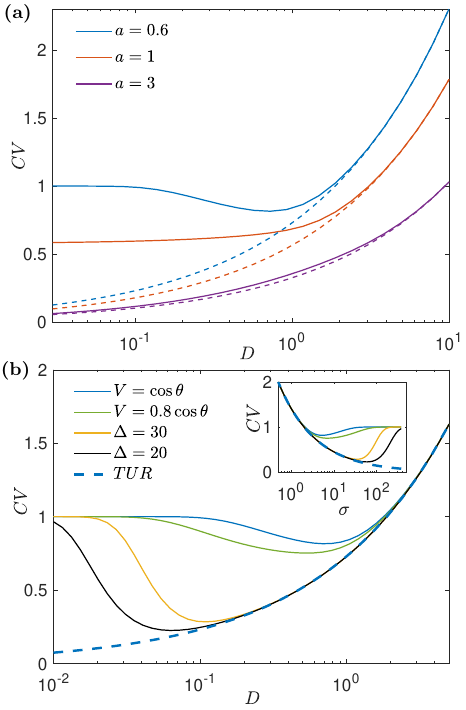}
\caption{(a) CV vs noise intensity for different excitability $a$ in a single excitable unit with $V(\theta)=\cos\theta$. The dashed line with the same color as the corresponding solid line is the lower bound predicted by the TUR. (b) CV vs noise intensity for different function $f$ with the same excitability $a=0.6$. The yellow and black solid lines correspond to the case $V(\theta)=\frac{\Delta}{\epsilon}\exp\left\{\epsilon(\cos\theta-1)\right\}$ with the same $\epsilon =100$ but different $\Delta$. The blue dashed line corresponds to the TUR bound \eqref{Eq:CV_bound}.}
\label{Fig:CoherenceRes}
\end{figure}

Coherence resonance \cite{pikovsky1997coherence,gang1993stochastic} is a phenomenon occurring widely in excitable systems, e.g. in various neuronal systems \cite{lindner2004effects,pisarchik2023coherence}, where there exists an optimal noise strength to make the oscillatory responses of the system most coherently. The coefficient of variance (CV), i.e. the square root of the fluctuation $\sqrt{F}$, is the common measurement of the coherence of the neuronal spikes. For the phase model \eqref{Eq:Model}, coherence resonance occurs when the system is in the excitable state with $a<1$. Here we choose $a=0.6$ as an example and calculate the CV of the ISI according to \eqref{Eq:I_pm} and \eqref{Eq:Fluc_ISI} numerically. A pronounced minimum in the dependence of CV versus noise strength is observed in Fig.~\ref{Fig:CoherenceRes} (a), meaning the optimal coherence happening here. Note that CV=1 when the noise is small corresponds to the Poisson process. 

For small noise intensity, the constraint due to SNIC bifurcation, i.e. the solid line with $a=1$, gives tighter general bound than that resulting from the TUR. However, for large noise intensity the TUR predicts a tighter bound instead. By direct substituting Eq.\eqref{Eq:EntropyProd} into Eq.\eqref{Eq:TUR}, we obtain the TUR bound exhibited in terms of the CV versus the noise intensity $D$ and excitability $a$
\begin{equation}
    CV \geq \sqrt{\frac{D}{a\pi}}.
    \label{Eq:CV_bound}
\end{equation}
What is new addition to the investigation of coherence resonance, comparing to a variety of previous studies in this field \cite{lindner2004effects,pisarchik2023coherence}, is that combining both dynamic and thermodynamic constraints on the coherence provides a more complete picture of the fluctuation. 
For the sake of completeness, we also investigate the super-threshold case with $a>1$.
The TUR captures well the lower bound of the CV and the SNIC bound with $a=1$ serves as a general upper bound for the superthreshold case, although it's not tight.

For different function $f(\theta)=-V'(\theta)$ with the same excitability $a$, the same TUR bound holds. As shown in Fig.~\ref{Fig:CoherenceRes} (b), we choose $V(\theta)=\cos\theta$, $0.8\cos\theta$ and a type-II-excitability-like potential \cite{lindner2001optimal} with $V(\theta)=\frac{\Delta}{\epsilon}\exp\left\{\epsilon(\cos\theta-1)\right\}$, respectively. The last type-II-excitability-like potential generates more coherent spike train than the $\cos\theta$ potential does and therefore and the minimum of fluctuation of the ISI becomes smaller than $\frac{1}{3}$.
We also plot the same data in dependence of the CV versus the entropy production $\sigma$ in the inset of Fig.~\ref{Fig:CoherenceRes} (b). We see that there exists
an optimal entropy production to maximize the coherence
. This can serve as a good example to show that more thermodynamic cost is necessary but not sufficient condition to achieve higher accuracy.

Above we only consider the case of a single unit, but in real excitable systems especially the neuronal systems, coupling effect of the neurons should not be ignored. So we consider the synchronization of coupled excitable units in the following section.

\section{Coherence-dissipation relation in the almost synchronous state}
Synchronization plays an important role in a variety of systems in physics, chemistry and biology \cite{pikovsky2001synchronization,kuramoto1984chemical}. In neuronal systems, synchronization are closely related to selective attention \cite{womelsdorf2007role}, memory formation \cite{fell2011role} and information processing \cite{singer1993synchronization} and so on. So considering constraints enforced by thermodynamics and dynamics in networks of neurons is at stake.
In this section, we consider $N$ active rotators coupled by a sine function of the phase differences:
\begin{equation}
    \dot{\theta}_i=\omega+\sin\theta_i+\frac{K}{N}\sum_{j=1}^{N}\sin(\theta_j-\theta_i)+\xi_i(t),
    \label{Eq:rotators_coupled}
\end{equation}
where $K$ is the coupling strength, $\xi(t)$ is the same Gaussian white noise as that of the single unit with strength $D$ and is independent for different oscillators. In the limiting case with no coupling, the coherence-dissipation relation should be the same as that of the single oscillator described in section \ref{Sec:single_oscillator}. We consider here the opposite limiting case with strong coupling so that all the oscillators are close to be in synchrony. Note that the system can not be in full synchrony due to the independent noise $\xi_i$, but could be close to be in synchrony if the coupling is strong enough.
In this almost synchronous state, $\theta_1\approx\theta_2\approx\cdots\approx\theta_N$, the $N$ degrees of freedom for the coupled equations \eqref{Eq:rotators_coupled} reduces to only that of the collective phase $\psi$ defined as $\psi_1=\frac{1}{N}\sum_i\theta_i$ as follows,
\begin{equation}
    \dot{\psi}_1 = \omega + \sin\psi_1 + \sqrt{\frac{1}{N}}\xi(t).
    \label{Eq:Model_coupled}
\end{equation}
The other $N-1$ degrees of freedom, which can be defined as $\psi_i = \theta_1-\theta_i$ for $i\in \{2, 3, \dots, N\}$, vanishes in the synchronous state.
The phase of the subsystem $\theta_1$ can be expressed as $\theta_1=\psi_1+(\psi_2+\cdots+\psi_N)/N$, so the dynamics of $\theta_1$ is described by Eq.~\eqref{Eq:Model_coupled} as $\theta_1\approx\psi_1$ in the almost synchronous state.
Note that Eq.~\eqref{Eq:Model_coupled} is the same as Eq.~\eqref{Eq:Model} by scaling the noise intensity by a factor $1/N$, so the thermodynamic cost is $1/N$ that of the uncoupled single case to achieve the same coherence. So the TUR for the subsystem in synchrony is given by
\begin{equation}
    F\geq \frac{2}{N\sigma},
\end{equation}
where $\sigma$ is the entropy production during mean time $\langle T\rangle$ for a single unit, i.e. Eq.~\eqref{Eq:EntropyProd}. 

\begin{figure}
\centering
\includegraphics{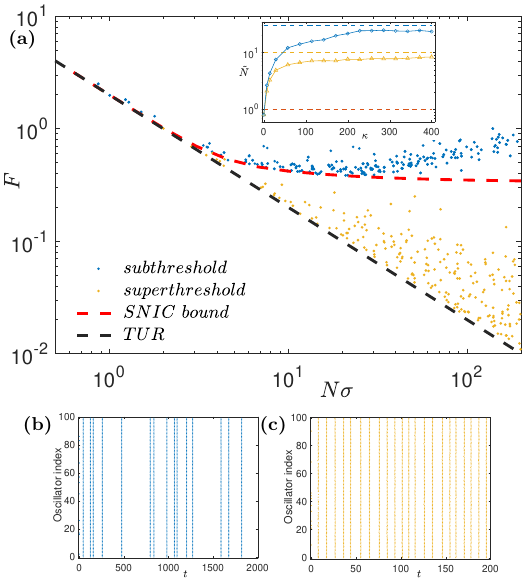}
\caption{(a) Fluctuation of the ISI $F$ from simulations vs the entropy production rate $\sigma$ for the strongly coupled excitable units with different number of oscillators $N$ both in the sub-threshold region (blue dots) and in the super-threshold region (orange dots). The red dashed line is the SNIC bound and the black dashed line is the TUR bound. Inset of (a) depicts how the quantity $\Tilde{N}$ varies with the normalized coupling strength $\kappa$ for $N=10$ (orange triangles) and $N=30$ (blue circles), respectively. The blue, orange and red dashed line corresponds to $\Tilde{N}=30, 10$ and $1$, respectively.
Raster plots of the coupled $100$ oscillators in the excitable region with $a=0.95$ (b) and oscillatory region with $a=1.2$ (c).}
\label{Fig:Fluc_vs_Nsigma}
\end{figure}

We directly simulate the coupled rotator model \eqref{Eq:rotators_coupled} for different number of oscillators $N$ and plot the fluctuation of the ISI versus $N$ multiplying the entropy production $\sigma$ of a single oscillator, as shown in Fig.~\ref{Fig:Fluc_vs_Nsigma} (a). To make the simulations more precise, we use the Euler-Maruyama method with small time interval $dt=0.0001$.
Similar to the single unit case, the excitable (subthreshold) and oscillatory (superthreshold) region are also separated by the SNIC bound represented by the red dashed line, as shown in Fig.~\ref{Fig:Fluc_vs_Nsigma} (a). Due to the strong coupling considered in this case, the excitable units are in almost synchronous state in both the subthreshold and superthreshold regions. We also check numerically how the thermodynamic bound varies with the coupling strength for small ensemble of oscillators with $N=10$ and $N=30$, respectively. As shown in the inset of Fig.~\ref{Fig:Fluc_vs_Nsigma} (a), the quantity $\Tilde{N}$ defined as $\Tilde{N}=\frac{2}{F\sigma}$ increases monotonically with the normalized coupling strength $\kappa$, which is defined as $\kappa=K/D$. In the case with no coupling, i.e. $\kappa=0$, the quantity $\Tilde{N}$ is bounded by $1$, while it is bounded by the number of coupling oscillators $N$ in the strong coupling case.
To visualize the spiking activity of the coupled excitable oscillators, we plot the spikes as the colored dots when the phase crosses a given value, i.e. using the raster plots commonly adopted for neurons. We simulate $100$ strongly coupled excitable units following Eq.~\eqref{Eq:rotators_coupled} and the raster plots in the excitable and oscillatory states are shown in Fig.~\ref{Fig:Fluc_vs_Nsigma} (b) and (c), respectively. We can see straightforwardly that the oscillators are almost in complete synchrony.
The only difference between Fig.~\ref{Fig:Fluc_vs_Nsigma} (a) and Fig.~\ref{Fig:TUR} (a) lies in that now the thermodynamics cost becomes $1/N$ that of the single unit. This is the advantage of synchronization, which could be used as a strategy to optimise the coherence of the system given a limited energy budget. 

\section{Conclusions}
Excitable systems such as neurons have distinctive coherence features compared with other oscillators having no excitability, e.g. the Stuart-Landau oscillators considered in previous studies \cite{cao2015free,fei2018design,lee2018thermodynamic,remlein2022coherence}.  
This is due to the constraints enforced not only by the TUR but also by the dynamical structure itself, such as the saddle-node structure in one dimensional excitable systems\cite{sigeti1989pseudo}. In the critical state, interesting diffusion phenomenon was observed in a recent experiment\cite{vsiler2018diffusing}.

In this paper, we combined the well-established results, i.e. the fluctuation of the ISI limited by $1/3$ and the coherence resonance phenomenon, together with the TUR developed in recent years to investigate the coherence in the excitable phase oscillators. We find quite different trade-off relation in the subthreshold (excitable) region and superthreshold (oscillatory) region, separated by the SNIC bound but meanwhile lower bounded by the TUR. Furthermore, we found that there is an optimal entropy production corresponding to the maximum coherence, which could serve as an alternative interpretation of the coherence resonance. It implies that more entropy production does not necessarily result in higher accuracy of currents. This can be intuitively understood such that the saddle-node structure induces a potential barrier to cross and therefore more payment of the thermodynamic cost does not necessarily transform effectively into oscillation. 

Coherence resonance has been studied in a variety of excitable systems including the more realistic neuron models \cite{lindner2004effects, pisarchik2023coherence}, e.g. the FitzHugh–Nagumo model and the Hodgkin-Huxley model, beyond the one-dimensional phase model considered in this paper. Nevertheless, we believe that the present results would shed new light on those previously studied cases. After all, considering both the dynamical and thermodynamic constraints is necessary for the inference of coherence in excitable systems. 

Lastly, we show that synchronization of more excitable oscillators favors coherence with lower thermodynamics cost. This may be related to the optimal information processing with constrained energy budget in real neuronal systems \cite{friston2010free,harris2012synaptic} and we leave it for future studies.

\section{Acknowledgements}
We acknowledge the supports from Yunnan Fundamental Research Projects (grant NO. 202401AU070216) and Scientific Research Fund Project (grant No. 2024J0012) of Education Department of Yunnan.

\normalem
\bibliography{references}

\begin{thebibliography}{55}%
\makeatletter
\providecommand \@ifxundefined [1]{%
 \@ifx{#1\undefined}
}%
\providecommand \@ifnum [1]{%
 \ifnum #1\expandafter \@firstoftwo
 \else \expandafter \@secondoftwo
 \fi
}%
\providecommand \@ifx [1]{%
 \ifx #1\expandafter \@firstoftwo
 \else \expandafter \@secondoftwo
 \fi
}%
\providecommand \natexlab [1]{#1}%
\providecommand \enquote  [1]{``#1''}%
\providecommand \bibnamefont  [1]{#1}%
\providecommand \bibfnamefont [1]{#1}%
\providecommand \citenamefont [1]{#1}%
\providecommand \href@noop [0]{\@secondoftwo}%
\providecommand \href [0]{\begingroup \@sanitize@url \@href}%
\providecommand \@href[1]{\@@startlink{#1}\@@href}%
\providecommand \@@href[1]{\endgroup#1\@@endlink}%
\providecommand \@sanitize@url [0]{\catcode `\\12\catcode `\$12\catcode `\&12\catcode `\#12\catcode `\^12\catcode `\_12\catcode `\%12\relax}%
\providecommand \@@startlink[1]{}%
\providecommand \@@endlink[0]{}%
\providecommand \url  [0]{\begingroup\@sanitize@url \@url }%
\providecommand \@url [1]{\endgroup\@href {#1}{\urlprefix }}%
\providecommand \urlprefix  [0]{URL }%
\providecommand \Eprint [0]{\href }%
\providecommand \doibase [0]{https://doi.org/}%
\providecommand \selectlanguage [0]{\@gobble}%
\providecommand \bibinfo  [0]{\@secondoftwo}%
\providecommand \bibfield  [0]{\@secondoftwo}%
\providecommand \translation [1]{[#1]}%
\providecommand \BibitemOpen [0]{}%
\providecommand \bibitemStop [0]{}%
\providecommand \bibitemNoStop [0]{.\EOS\space}%
\providecommand \EOS [0]{\spacefactor3000\relax}%
\providecommand \BibitemShut  [1]{\csname bibitem#1\endcsname}%
\let\auto@bib@innerbib\@empty
\bibitem [{\citenamefont {Barato}\ and\ \citenamefont {Seifert}(2015)}]{barato2015thermodynamic}%
  \BibitemOpen
  \bibfield  {author} {\bibinfo {author} {\bibfnamefont {A.~C.}\ \bibnamefont {Barato}}\ and\ \bibinfo {author} {\bibfnamefont {U.}~\bibnamefont {Seifert}},\ }\bibfield  {title} {\bibinfo {title} {Thermodynamic uncertainty relation for biomolecular processes},\ }\href@noop {} {\bibfield  {journal} {\bibinfo  {journal} {Physical review letters}\ }\textbf {\bibinfo {volume} {114}},\ \bibinfo {pages} {158101} (\bibinfo {year} {2015})}\BibitemShut {NoStop}%
\bibitem [{\citenamefont {Barato}\ and\ \citenamefont {Seifert}(2016)}]{barato2016cost}%
  \BibitemOpen
  \bibfield  {author} {\bibinfo {author} {\bibfnamefont {A.~C.}\ \bibnamefont {Barato}}\ and\ \bibinfo {author} {\bibfnamefont {U.}~\bibnamefont {Seifert}},\ }\bibfield  {title} {\bibinfo {title} {Cost and precision of brownian clocks},\ }\href@noop {} {\bibfield  {journal} {\bibinfo  {journal} {Physical Review X}\ }\textbf {\bibinfo {volume} {6}},\ \bibinfo {pages} {041053} (\bibinfo {year} {2016})}\BibitemShut {NoStop}%
\bibitem [{\citenamefont {Gingrich}\ \emph {et~al.}(2016)\citenamefont {Gingrich}, \citenamefont {Horowitz}, \citenamefont {Perunov},\ and\ \citenamefont {England}}]{gingrich2016dissipation}%
  \BibitemOpen
  \bibfield  {author} {\bibinfo {author} {\bibfnamefont {T.~R.}\ \bibnamefont {Gingrich}}, \bibinfo {author} {\bibfnamefont {J.~M.}\ \bibnamefont {Horowitz}}, \bibinfo {author} {\bibfnamefont {N.}~\bibnamefont {Perunov}},\ and\ \bibinfo {author} {\bibfnamefont {J.~L.}\ \bibnamefont {England}},\ }\bibfield  {title} {\bibinfo {title} {Dissipation bounds all steady-state current fluctuations},\ }\href@noop {} {\bibfield  {journal} {\bibinfo  {journal} {Physical review letters}\ }\textbf {\bibinfo {volume} {116}},\ \bibinfo {pages} {120601} (\bibinfo {year} {2016})}\BibitemShut {NoStop}%
\bibitem [{\citenamefont {Pietzonka}\ \emph {et~al.}(2016)\citenamefont {Pietzonka}, \citenamefont {Barato},\ and\ \citenamefont {Seifert}}]{pietzonka2016universal}%
  \BibitemOpen
  \bibfield  {author} {\bibinfo {author} {\bibfnamefont {P.}~\bibnamefont {Pietzonka}}, \bibinfo {author} {\bibfnamefont {A.~C.}\ \bibnamefont {Barato}},\ and\ \bibinfo {author} {\bibfnamefont {U.}~\bibnamefont {Seifert}},\ }\bibfield  {title} {\bibinfo {title} {Universal bounds on current fluctuations},\ }\href@noop {} {\bibfield  {journal} {\bibinfo  {journal} {Physical Review E}\ }\textbf {\bibinfo {volume} {93}},\ \bibinfo {pages} {052145} (\bibinfo {year} {2016})}\BibitemShut {NoStop}%
\bibitem [{\citenamefont {Garrahan}(2017)}]{garrahan2017simple}%
  \BibitemOpen
  \bibfield  {author} {\bibinfo {author} {\bibfnamefont {J.~P.}\ \bibnamefont {Garrahan}},\ }\bibfield  {title} {\bibinfo {title} {Simple bounds on fluctuations and uncertainty relations for first-passage times of counting observables},\ }\href@noop {} {\bibfield  {journal} {\bibinfo  {journal} {Physical Review E}\ }\textbf {\bibinfo {volume} {95}},\ \bibinfo {pages} {032134} (\bibinfo {year} {2017})}\BibitemShut {NoStop}%
\bibitem [{\citenamefont {Gingrich}\ and\ \citenamefont {Horowitz}(2017)}]{gingrich2017fundamental}%
  \BibitemOpen
  \bibfield  {author} {\bibinfo {author} {\bibfnamefont {T.~R.}\ \bibnamefont {Gingrich}}\ and\ \bibinfo {author} {\bibfnamefont {J.~M.}\ \bibnamefont {Horowitz}},\ }\bibfield  {title} {\bibinfo {title} {Fundamental bounds on first passage time fluctuations for currents},\ }\href@noop {} {\bibfield  {journal} {\bibinfo  {journal} {Physical review letters}\ }\textbf {\bibinfo {volume} {119}},\ \bibinfo {pages} {170601} (\bibinfo {year} {2017})}\BibitemShut {NoStop}%
\bibitem [{\citenamefont {Neri}\ \emph {et~al.}(2017)\citenamefont {Neri}, \citenamefont {Rold{\'a}n},\ and\ \citenamefont {J{\"u}licher}}]{neri2017statistics}%
  \BibitemOpen
  \bibfield  {author} {\bibinfo {author} {\bibfnamefont {I.}~\bibnamefont {Neri}}, \bibinfo {author} {\bibfnamefont {{\'E}.}~\bibnamefont {Rold{\'a}n}},\ and\ \bibinfo {author} {\bibfnamefont {F.}~\bibnamefont {J{\"u}licher}},\ }\bibfield  {title} {\bibinfo {title} {Statistics of infima and stopping times of entropy production and applications to active molecular processes},\ }\href@noop {} {\bibfield  {journal} {\bibinfo  {journal} {Physical Review X}\ }\textbf {\bibinfo {volume} {7}},\ \bibinfo {pages} {011019} (\bibinfo {year} {2017})}\BibitemShut {NoStop}%
\bibitem [{\citenamefont {Macieszczak}\ \emph {et~al.}(2018)\citenamefont {Macieszczak}, \citenamefont {Brandner},\ and\ \citenamefont {Garrahan}}]{macieszczak2018unified}%
  \BibitemOpen
  \bibfield  {author} {\bibinfo {author} {\bibfnamefont {K.}~\bibnamefont {Macieszczak}}, \bibinfo {author} {\bibfnamefont {K.}~\bibnamefont {Brandner}},\ and\ \bibinfo {author} {\bibfnamefont {J.~P.}\ \bibnamefont {Garrahan}},\ }\bibfield  {title} {\bibinfo {title} {Unified thermodynamic uncertainty relations in linear response},\ }\href@noop {} {\bibfield  {journal} {\bibinfo  {journal} {Physical review letters}\ }\textbf {\bibinfo {volume} {121}},\ \bibinfo {pages} {130601} (\bibinfo {year} {2018})}\BibitemShut {NoStop}%
\bibitem [{\citenamefont {Lee}\ \emph {et~al.}(2018)\citenamefont {Lee}, \citenamefont {Hyeon},\ and\ \citenamefont {Jo}}]{lee2018thermodynamic}%
  \BibitemOpen
  \bibfield  {author} {\bibinfo {author} {\bibfnamefont {S.}~\bibnamefont {Lee}}, \bibinfo {author} {\bibfnamefont {C.}~\bibnamefont {Hyeon}},\ and\ \bibinfo {author} {\bibfnamefont {J.}~\bibnamefont {Jo}},\ }\bibfield  {title} {\bibinfo {title} {Thermodynamic uncertainty relation of interacting oscillators in synchrony},\ }\href@noop {} {\bibfield  {journal} {\bibinfo  {journal} {Physical Review E}\ }\textbf {\bibinfo {volume} {98}},\ \bibinfo {pages} {032119} (\bibinfo {year} {2018})}\BibitemShut {NoStop}%
\bibitem [{\citenamefont {Hasegawa}\ and\ \citenamefont {Van~Vu}(2019)}]{hasegawa2019fluctuation}%
  \BibitemOpen
  \bibfield  {author} {\bibinfo {author} {\bibfnamefont {Y.}~\bibnamefont {Hasegawa}}\ and\ \bibinfo {author} {\bibfnamefont {T.}~\bibnamefont {Van~Vu}},\ }\bibfield  {title} {\bibinfo {title} {Fluctuation theorem uncertainty relation},\ }\href@noop {} {\bibfield  {journal} {\bibinfo  {journal} {Physical review letters}\ }\textbf {\bibinfo {volume} {123}},\ \bibinfo {pages} {110602} (\bibinfo {year} {2019})}\BibitemShut {NoStop}%
\bibitem [{\citenamefont {Hasegawa}(2020)}]{hasegawa2020quantum}%
  \BibitemOpen
  \bibfield  {author} {\bibinfo {author} {\bibfnamefont {Y.}~\bibnamefont {Hasegawa}},\ }\bibfield  {title} {\bibinfo {title} {Quantum thermodynamic uncertainty relation for continuous measurement},\ }\href@noop {} {\bibfield  {journal} {\bibinfo  {journal} {Physical Review Letters}\ }\textbf {\bibinfo {volume} {125}},\ \bibinfo {pages} {050601} (\bibinfo {year} {2020})}\BibitemShut {NoStop}%
\bibitem [{\citenamefont {Liu}\ \emph {et~al.}(2020)\citenamefont {Liu}, \citenamefont {Gong},\ and\ \citenamefont {Ueda}}]{liu2020thermodynamic}%
  \BibitemOpen
  \bibfield  {author} {\bibinfo {author} {\bibfnamefont {K.}~\bibnamefont {Liu}}, \bibinfo {author} {\bibfnamefont {Z.}~\bibnamefont {Gong}},\ and\ \bibinfo {author} {\bibfnamefont {M.}~\bibnamefont {Ueda}},\ }\bibfield  {title} {\bibinfo {title} {Thermodynamic uncertainty relation for arbitrary initial states},\ }\href@noop {} {\bibfield  {journal} {\bibinfo  {journal} {Physical Review Letters}\ }\textbf {\bibinfo {volume} {125}},\ \bibinfo {pages} {140602} (\bibinfo {year} {2020})}\BibitemShut {NoStop}%
\bibitem [{\citenamefont {Falasco}\ \emph {et~al.}(2020)\citenamefont {Falasco}, \citenamefont {Esposito},\ and\ \citenamefont {Delvenne}}]{falasco2020unifying}%
  \BibitemOpen
  \bibfield  {author} {\bibinfo {author} {\bibfnamefont {G.}~\bibnamefont {Falasco}}, \bibinfo {author} {\bibfnamefont {M.}~\bibnamefont {Esposito}},\ and\ \bibinfo {author} {\bibfnamefont {J.-C.}\ \bibnamefont {Delvenne}},\ }\bibfield  {title} {\bibinfo {title} {Unifying thermodynamic uncertainty relations},\ }\href@noop {} {\bibfield  {journal} {\bibinfo  {journal} {New Journal of Physics}\ }\textbf {\bibinfo {volume} {22}},\ \bibinfo {pages} {053046} (\bibinfo {year} {2020})}\BibitemShut {NoStop}%
\bibitem [{\citenamefont {Pearson}\ \emph {et~al.}(2021)\citenamefont {Pearson}, \citenamefont {Guryanova}, \citenamefont {Erker}, \citenamefont {Laird}, \citenamefont {Briggs}, \citenamefont {Huber},\ and\ \citenamefont {Ares}}]{pearson2021measuring}%
  \BibitemOpen
  \bibfield  {author} {\bibinfo {author} {\bibfnamefont {A.~N.}\ \bibnamefont {Pearson}}, \bibinfo {author} {\bibfnamefont {Y.}~\bibnamefont {Guryanova}}, \bibinfo {author} {\bibfnamefont {P.}~\bibnamefont {Erker}}, \bibinfo {author} {\bibfnamefont {E.~A.}\ \bibnamefont {Laird}}, \bibinfo {author} {\bibfnamefont {G.~A.~D.}\ \bibnamefont {Briggs}}, \bibinfo {author} {\bibfnamefont {M.}~\bibnamefont {Huber}},\ and\ \bibinfo {author} {\bibfnamefont {N.}~\bibnamefont {Ares}},\ }\bibfield  {title} {\bibinfo {title} {Measuring the thermodynamic cost of timekeeping},\ }\href@noop {} {\bibfield  {journal} {\bibinfo  {journal} {Physical Review X}\ }\textbf {\bibinfo {volume} {11}},\ \bibinfo {pages} {021029} (\bibinfo {year} {2021})}\BibitemShut {NoStop}%
\bibitem [{\citenamefont {Dechant}\ and\ \citenamefont {Sasa}(2021)}]{dechant2021continuous}%
  \BibitemOpen
  \bibfield  {author} {\bibinfo {author} {\bibfnamefont {A.}~\bibnamefont {Dechant}}\ and\ \bibinfo {author} {\bibfnamefont {S.-i.}\ \bibnamefont {Sasa}},\ }\bibfield  {title} {\bibinfo {title} {Continuous time reversal and equality in the thermodynamic uncertainty relation},\ }\href@noop {} {\bibfield  {journal} {\bibinfo  {journal} {Physical Review Research}\ }\textbf {\bibinfo {volume} {3}},\ \bibinfo {pages} {L042012} (\bibinfo {year} {2021})}\BibitemShut {NoStop}%
\bibitem [{\citenamefont {Hasegawa}(2021)}]{hasegawa2021thermodynamic}%
  \BibitemOpen
  \bibfield  {author} {\bibinfo {author} {\bibfnamefont {Y.}~\bibnamefont {Hasegawa}},\ }\bibfield  {title} {\bibinfo {title} {Thermodynamic uncertainty relation for general open quantum systems},\ }\href@noop {} {\bibfield  {journal} {\bibinfo  {journal} {Physical Review Letters}\ }\textbf {\bibinfo {volume} {126}},\ \bibinfo {pages} {010602} (\bibinfo {year} {2021})}\BibitemShut {NoStop}%
\bibitem [{\citenamefont {Yoshimura}\ and\ \citenamefont {Ito}(2021)}]{yoshimura2021thermodynamic}%
  \BibitemOpen
  \bibfield  {author} {\bibinfo {author} {\bibfnamefont {K.}~\bibnamefont {Yoshimura}}\ and\ \bibinfo {author} {\bibfnamefont {S.}~\bibnamefont {Ito}},\ }\bibfield  {title} {\bibinfo {title} {Thermodynamic uncertainty relation and thermodynamic speed limit in deterministic chemical reaction networks},\ }\href@noop {} {\bibfield  {journal} {\bibinfo  {journal} {Physical review letters}\ }\textbf {\bibinfo {volume} {127}},\ \bibinfo {pages} {160601} (\bibinfo {year} {2021})}\BibitemShut {NoStop}%
\bibitem [{\citenamefont {Pietzonka}(2022)}]{pietzonka2022classical}%
  \BibitemOpen
  \bibfield  {author} {\bibinfo {author} {\bibfnamefont {P.}~\bibnamefont {Pietzonka}},\ }\bibfield  {title} {\bibinfo {title} {Classical pendulum clocks break the thermodynamic uncertainty relation},\ }\href@noop {} {\bibfield  {journal} {\bibinfo  {journal} {Physical Review Letters}\ }\textbf {\bibinfo {volume} {128}},\ \bibinfo {pages} {130606} (\bibinfo {year} {2022})}\BibitemShut {NoStop}%
\bibitem [{\citenamefont {Van~Vu}\ and\ \citenamefont {Saito}(2023)}]{van2023thermodynamic}%
  \BibitemOpen
  \bibfield  {author} {\bibinfo {author} {\bibfnamefont {T.}~\bibnamefont {Van~Vu}}\ and\ \bibinfo {author} {\bibfnamefont {K.}~\bibnamefont {Saito}},\ }\bibfield  {title} {\bibinfo {title} {Thermodynamic unification of optimal transport: Thermodynamic uncertainty relation, minimum dissipation, and thermodynamic speed limits},\ }\href@noop {} {\bibfield  {journal} {\bibinfo  {journal} {Physical Review X}\ }\textbf {\bibinfo {volume} {13}},\ \bibinfo {pages} {011013} (\bibinfo {year} {2023})}\BibitemShut {NoStop}%
\bibitem [{\citenamefont {Dieball}\ and\ \citenamefont {Godec}(2023)}]{dieball2023direct}%
  \BibitemOpen
  \bibfield  {author} {\bibinfo {author} {\bibfnamefont {C.}~\bibnamefont {Dieball}}\ and\ \bibinfo {author} {\bibfnamefont {A.}~\bibnamefont {Godec}},\ }\bibfield  {title} {\bibinfo {title} {Direct route to thermodynamic uncertainty relations and their saturation},\ }\href@noop {} {\bibfield  {journal} {\bibinfo  {journal} {Physical Review Letters}\ }\textbf {\bibinfo {volume} {130}},\ \bibinfo {pages} {087101} (\bibinfo {year} {2023})}\BibitemShut {NoStop}%
\bibitem [{\citenamefont {Pietzonka}\ and\ \citenamefont {Coghi}(2024)}]{pietzonka2024thermodynamic}%
  \BibitemOpen
  \bibfield  {author} {\bibinfo {author} {\bibfnamefont {P.}~\bibnamefont {Pietzonka}}\ and\ \bibinfo {author} {\bibfnamefont {F.}~\bibnamefont {Coghi}},\ }\bibfield  {title} {\bibinfo {title} {Thermodynamic cost for precision of general counting observables},\ }\href@noop {} {\bibfield  {journal} {\bibinfo  {journal} {Physical Review E}\ }\textbf {\bibinfo {volume} {109}},\ \bibinfo {pages} {064128} (\bibinfo {year} {2024})}\BibitemShut {NoStop}%
\bibitem [{\citenamefont {Dechant}\ and\ \citenamefont {Sasa}(2018)}]{dechant2018current}%
  \BibitemOpen
  \bibfield  {author} {\bibinfo {author} {\bibfnamefont {A.}~\bibnamefont {Dechant}}\ and\ \bibinfo {author} {\bibfnamefont {S.-i.}\ \bibnamefont {Sasa}},\ }\bibfield  {title} {\bibinfo {title} {Current fluctuations and transport efficiency for general langevin systems},\ }\href@noop {} {\bibfield  {journal} {\bibinfo  {journal} {Journal of Statistical Mechanics: Theory and Experiment}\ }\textbf {\bibinfo {volume} {2018}},\ \bibinfo {pages} {063209} (\bibinfo {year} {2018})}\BibitemShut {NoStop}%
\bibitem [{\citenamefont {Cao}\ \emph {et~al.}(2015)\citenamefont {Cao}, \citenamefont {Wang}, \citenamefont {Ouyang},\ and\ \citenamefont {Tu}}]{cao2015free}%
  \BibitemOpen
  \bibfield  {author} {\bibinfo {author} {\bibfnamefont {Y.}~\bibnamefont {Cao}}, \bibinfo {author} {\bibfnamefont {H.}~\bibnamefont {Wang}}, \bibinfo {author} {\bibfnamefont {Q.}~\bibnamefont {Ouyang}},\ and\ \bibinfo {author} {\bibfnamefont {Y.}~\bibnamefont {Tu}},\ }\bibfield  {title} {\bibinfo {title} {The free-energy cost of accurate biochemical oscillations},\ }\href@noop {} {\bibfield  {journal} {\bibinfo  {journal} {Nature physics}\ }\textbf {\bibinfo {volume} {11}},\ \bibinfo {pages} {772} (\bibinfo {year} {2015})}\BibitemShut {NoStop}%
\bibitem [{\citenamefont {Fei}\ \emph {et~al.}(2018)\citenamefont {Fei}, \citenamefont {Cao}, \citenamefont {Ouyang},\ and\ \citenamefont {Tu}}]{fei2018design}%
  \BibitemOpen
  \bibfield  {author} {\bibinfo {author} {\bibfnamefont {C.}~\bibnamefont {Fei}}, \bibinfo {author} {\bibfnamefont {Y.}~\bibnamefont {Cao}}, \bibinfo {author} {\bibfnamefont {Q.}~\bibnamefont {Ouyang}},\ and\ \bibinfo {author} {\bibfnamefont {Y.}~\bibnamefont {Tu}},\ }\bibfield  {title} {\bibinfo {title} {Design principles for enhancing phase sensitivity and suppressing phase fluctuations simultaneously in biochemical oscillatory systems},\ }\href@noop {} {\bibfield  {journal} {\bibinfo  {journal} {Nature communications}\ }\textbf {\bibinfo {volume} {9}},\ \bibinfo {pages} {1434} (\bibinfo {year} {2018})}\BibitemShut {NoStop}%
\bibitem [{\citenamefont {Izhikevich}(2007)}]{izhikevich2007dynamical}%
  \BibitemOpen
  \bibfield  {author} {\bibinfo {author} {\bibfnamefont {E.~M.}\ \bibnamefont {Izhikevich}},\ }\href@noop {} {\emph {\bibinfo {title} {Dynamical systems in neuroscience}}}\ (\bibinfo  {publisher} {MIT press},\ \bibinfo {year} {2007})\BibitemShut {NoStop}%
\bibitem [{\citenamefont {Lindner}\ \emph {et~al.}(2003)\citenamefont {Lindner}, \citenamefont {Longtin},\ and\ \citenamefont {Bulsara}}]{lindner2003analytic}%
  \BibitemOpen
  \bibfield  {author} {\bibinfo {author} {\bibfnamefont {B.}~\bibnamefont {Lindner}}, \bibinfo {author} {\bibfnamefont {A.}~\bibnamefont {Longtin}},\ and\ \bibinfo {author} {\bibfnamefont {A.}~\bibnamefont {Bulsara}},\ }\bibfield  {title} {\bibinfo {title} {Analytic expressions for rate and cv of a type i neuron driven by white gaussian noise},\ }\href@noop {} {\bibfield  {journal} {\bibinfo  {journal} {Neural computation}\ }\textbf {\bibinfo {volume} {15}},\ \bibinfo {pages} {1761} (\bibinfo {year} {2003})}\BibitemShut {NoStop}%
\bibitem [{\citenamefont {Ermentrout}\ and\ \citenamefont {Kopell}(1986)}]{ermentrout1986parabolic}%
  \BibitemOpen
  \bibfield  {author} {\bibinfo {author} {\bibfnamefont {G.~B.}\ \bibnamefont {Ermentrout}}\ and\ \bibinfo {author} {\bibfnamefont {N.}~\bibnamefont {Kopell}},\ }\bibfield  {title} {\bibinfo {title} {Parabolic bursting in an excitable system coupled with a slow oscillation},\ }\href@noop {} {\bibfield  {journal} {\bibinfo  {journal} {SIAM journal on applied mathematics}\ }\textbf {\bibinfo {volume} {46}},\ \bibinfo {pages} {233} (\bibinfo {year} {1986})}\BibitemShut {NoStop}%
\bibitem [{\citenamefont {Sigeti}\ and\ \citenamefont {Horsthemke}(1989)}]{sigeti1989pseudo}%
  \BibitemOpen
  \bibfield  {author} {\bibinfo {author} {\bibfnamefont {D.}~\bibnamefont {Sigeti}}\ and\ \bibinfo {author} {\bibfnamefont {W.}~\bibnamefont {Horsthemke}},\ }\bibfield  {title} {\bibinfo {title} {Pseudo-regular oscillations induced by external noise},\ }\href@noop {} {\bibfield  {journal} {\bibinfo  {journal} {Journal of statistical physics}\ }\textbf {\bibinfo {volume} {54}},\ \bibinfo {pages} {1217} (\bibinfo {year} {1989})}\BibitemShut {NoStop}%
\bibitem [{\citenamefont {Lindner}\ \emph {et~al.}(2004)\citenamefont {Lindner}, \citenamefont {Garc{\i}a-Ojalvo}, \citenamefont {Neiman},\ and\ \citenamefont {Schimansky-Geier}}]{lindner2004effects}%
  \BibitemOpen
  \bibfield  {author} {\bibinfo {author} {\bibfnamefont {B.}~\bibnamefont {Lindner}}, \bibinfo {author} {\bibfnamefont {J.}~\bibnamefont {Garc{\i}a-Ojalvo}}, \bibinfo {author} {\bibfnamefont {A.}~\bibnamefont {Neiman}},\ and\ \bibinfo {author} {\bibfnamefont {L.}~\bibnamefont {Schimansky-Geier}},\ }\bibfield  {title} {\bibinfo {title} {Effects of noise in excitable systems},\ }\href@noop {} {\bibfield  {journal} {\bibinfo  {journal} {Physics reports}\ }\textbf {\bibinfo {volume} {392}},\ \bibinfo {pages} {321} (\bibinfo {year} {2004})}\BibitemShut {NoStop}%
\bibitem [{\citenamefont {Pikovsky}\ and\ \citenamefont {Kurths}(1997)}]{pikovsky1997coherence}%
  \BibitemOpen
  \bibfield  {author} {\bibinfo {author} {\bibfnamefont {A.~S.}\ \bibnamefont {Pikovsky}}\ and\ \bibinfo {author} {\bibfnamefont {J.}~\bibnamefont {Kurths}},\ }\bibfield  {title} {\bibinfo {title} {Coherence resonance in a noise-driven excitable system},\ }\href@noop {} {\bibfield  {journal} {\bibinfo  {journal} {Physical Review Letters}\ }\textbf {\bibinfo {volume} {78}},\ \bibinfo {pages} {775} (\bibinfo {year} {1997})}\BibitemShut {NoStop}%
\bibitem [{\citenamefont {Kiss}\ \emph {et~al.}(2003)\citenamefont {Kiss}, \citenamefont {Hudson}, \citenamefont {Escalera~Santos},\ and\ \citenamefont {Parmananda}}]{kiss2003experiments}%
  \BibitemOpen
  \bibfield  {author} {\bibinfo {author} {\bibfnamefont {I.~Z.}\ \bibnamefont {Kiss}}, \bibinfo {author} {\bibfnamefont {J.~L.}\ \bibnamefont {Hudson}}, \bibinfo {author} {\bibfnamefont {G.~J.}\ \bibnamefont {Escalera~Santos}},\ and\ \bibinfo {author} {\bibfnamefont {P.}~\bibnamefont {Parmananda}},\ }\bibfield  {title} {\bibinfo {title} {Experiments on coherence resonance: Noisy precursors to hopf bifurcations},\ }\href@noop {} {\bibfield  {journal} {\bibinfo  {journal} {Physical Review E}\ }\textbf {\bibinfo {volume} {67}},\ \bibinfo {pages} {035201(R)} (\bibinfo {year} {2003})}\BibitemShut {NoStop}%
\bibitem [{\citenamefont {Ushakov}\ \emph {et~al.}(2005)\citenamefont {Ushakov}, \citenamefont {W\"unsche}, \citenamefont {Henneberger}, \citenamefont {Khovanov}, \citenamefont {Schimansky-Geier},\ and\ \citenamefont {Zaks}}]{ushakov2005coherence}%
  \BibitemOpen
  \bibfield  {author} {\bibinfo {author} {\bibfnamefont {O.~V.}\ \bibnamefont {Ushakov}}, \bibinfo {author} {\bibfnamefont {H.-J.}\ \bibnamefont {W\"unsche}}, \bibinfo {author} {\bibfnamefont {F.}~\bibnamefont {Henneberger}}, \bibinfo {author} {\bibfnamefont {I.~A.}\ \bibnamefont {Khovanov}}, \bibinfo {author} {\bibfnamefont {L.}~\bibnamefont {Schimansky-Geier}},\ and\ \bibinfo {author} {\bibfnamefont {M.~A.}\ \bibnamefont {Zaks}},\ }\bibfield  {title} {\bibinfo {title} {Coherence resonance near a hopf bifurcation},\ }\href@noop {} {\bibfield  {journal} {\bibinfo  {journal} {Physical review letters}\ }\textbf {\bibinfo {volume} {95}},\ \bibinfo {pages} {123903} (\bibinfo {year} {2005})}\BibitemShut {NoStop}%
\bibitem [{\citenamefont {Lee}\ \emph {et~al.}(2010)\citenamefont {Lee}, \citenamefont {Choi}, \citenamefont {Han},\ and\ \citenamefont {Strano}}]{lee2010coherence}%
  \BibitemOpen
  \bibfield  {author} {\bibinfo {author} {\bibfnamefont {C.~Y.}\ \bibnamefont {Lee}}, \bibinfo {author} {\bibfnamefont {W.}~\bibnamefont {Choi}}, \bibinfo {author} {\bibfnamefont {J.-H.}\ \bibnamefont {Han}},\ and\ \bibinfo {author} {\bibfnamefont {M.~S.}\ \bibnamefont {Strano}},\ }\bibfield  {title} {\bibinfo {title} {Coherence resonance in a single-walled carbon nanotube ion channel},\ }\href@noop {} {\bibfield  {journal} {\bibinfo  {journal} {Science}\ }\textbf {\bibinfo {volume} {329}},\ \bibinfo {pages} {1320} (\bibinfo {year} {2010})}\BibitemShut {NoStop}%
\bibitem [{\citenamefont {Mompo}\ \emph {et~al.}(2018)\citenamefont {Mompo}, \citenamefont {Ruiz-Garcia}, \citenamefont {Carretero}, \citenamefont {Grahn}, \citenamefont {Zhang},\ and\ \citenamefont {Bonilla}}]{mompo2018coherence}%
  \BibitemOpen
  \bibfield  {author} {\bibinfo {author} {\bibfnamefont {E.}~\bibnamefont {Mompo}}, \bibinfo {author} {\bibfnamefont {M.}~\bibnamefont {Ruiz-Garcia}}, \bibinfo {author} {\bibfnamefont {M.}~\bibnamefont {Carretero}}, \bibinfo {author} {\bibfnamefont {H.~T.}\ \bibnamefont {Grahn}}, \bibinfo {author} {\bibfnamefont {Y.}~\bibnamefont {Zhang}},\ and\ \bibinfo {author} {\bibfnamefont {L.~L.}\ \bibnamefont {Bonilla}},\ }\bibfield  {title} {\bibinfo {title} {Coherence resonance and stochastic resonance in an excitable semiconductor superlattice},\ }\href@noop {} {\bibfield  {journal} {\bibinfo  {journal} {Physical Review Letters}\ }\textbf {\bibinfo {volume} {121}},\ \bibinfo {pages} {086805} (\bibinfo {year} {2018})}\BibitemShut {NoStop}%
\bibitem [{\citenamefont {Kromer}\ \emph {et~al.}(2016)\citenamefont {Kromer}, \citenamefont {Schimansky-Geier},\ and\ \citenamefont {Neiman}}]{kromer2016emergence}%
  \BibitemOpen
  \bibfield  {author} {\bibinfo {author} {\bibfnamefont {J.~A.}\ \bibnamefont {Kromer}}, \bibinfo {author} {\bibfnamefont {L.}~\bibnamefont {Schimansky-Geier}},\ and\ \bibinfo {author} {\bibfnamefont {A.~B.}\ \bibnamefont {Neiman}},\ }\bibfield  {title} {\bibinfo {title} {Emergence and coherence of oscillations in star networks of stochastic excitable elements},\ }\href@noop {} {\bibfield  {journal} {\bibinfo  {journal} {Physical Review E}\ }\textbf {\bibinfo {volume} {93}},\ \bibinfo {pages} {042406} (\bibinfo {year} {2016})}\BibitemShut {NoStop}%
\bibitem [{\citenamefont {Zheng}\ and\ \citenamefont {Pikovsky}(2018)}]{zheng2018delay}%
  \BibitemOpen
  \bibfield  {author} {\bibinfo {author} {\bibfnamefont {C.}~\bibnamefont {Zheng}}\ and\ \bibinfo {author} {\bibfnamefont {A.}~\bibnamefont {Pikovsky}},\ }\bibfield  {title} {\bibinfo {title} {Delay-induced stochastic bursting in excitable noisy systems},\ }\href@noop {} {\bibfield  {journal} {\bibinfo  {journal} {Physical Review E}\ }\textbf {\bibinfo {volume} {98}},\ \bibinfo {pages} {042148} (\bibinfo {year} {2018})}\BibitemShut {NoStop}%
\bibitem [{\citenamefont {Buend{\'\i}a}\ \emph {et~al.}(2021)\citenamefont {Buend{\'\i}a}, \citenamefont {Villegas}, \citenamefont {Burioni},\ and\ \citenamefont {Mu{\~n}oz}}]{buendia2021hybrid}%
  \BibitemOpen
  \bibfield  {author} {\bibinfo {author} {\bibfnamefont {V.}~\bibnamefont {Buend{\'\i}a}}, \bibinfo {author} {\bibfnamefont {P.}~\bibnamefont {Villegas}}, \bibinfo {author} {\bibfnamefont {R.}~\bibnamefont {Burioni}},\ and\ \bibinfo {author} {\bibfnamefont {M.~A.}\ \bibnamefont {Mu{\~n}oz}},\ }\bibfield  {title} {\bibinfo {title} {Hybrid-type synchronization transitions: Where incipient oscillations, scale-free avalanches, and bistability live together},\ }\href@noop {} {\bibfield  {journal} {\bibinfo  {journal} {Physical Review Research}\ }\textbf {\bibinfo {volume} {3}},\ \bibinfo {pages} {023224} (\bibinfo {year} {2021})}\BibitemShut {NoStop}%
\bibitem [{\citenamefont {Seifert}(2012)}]{seifert2012stochastic}%
  \BibitemOpen
  \bibfield  {author} {\bibinfo {author} {\bibfnamefont {U.}~\bibnamefont {Seifert}},\ }\bibfield  {title} {\bibinfo {title} {Stochastic thermodynamics, fluctuation theorems and molecular machines},\ }\href@noop {} {\bibfield  {journal} {\bibinfo  {journal} {Reports on progress in physics}\ }\textbf {\bibinfo {volume} {75}},\ \bibinfo {pages} {126001} (\bibinfo {year} {2012})}\BibitemShut {NoStop}%
\bibitem [{\citenamefont {Risken}(1996)}]{risken1996fokker}%
  \BibitemOpen
  \bibfield  {author} {\bibinfo {author} {\bibfnamefont {H.}~\bibnamefont {Risken}},\ }\href@noop {} {\bibinfo {title} {The fokker-planck equation}} (\bibinfo {year} {1996})\BibitemShut {NoStop}%
\bibitem [{\citenamefont {Reimann}\ \emph {et~al.}(2001)\citenamefont {Reimann}, \citenamefont {Van~den Broeck}, \citenamefont {Linke}, \citenamefont {H\"anggi}, \citenamefont {Rubi},\ and\ \citenamefont {P\'erez-Madrid}}]{reimann2001giant}%
  \BibitemOpen
  \bibfield  {author} {\bibinfo {author} {\bibfnamefont {P.}~\bibnamefont {Reimann}}, \bibinfo {author} {\bibfnamefont {C.}~\bibnamefont {Van~den Broeck}}, \bibinfo {author} {\bibfnamefont {H.}~\bibnamefont {Linke}}, \bibinfo {author} {\bibfnamefont {P.}~\bibnamefont {H\"anggi}}, \bibinfo {author} {\bibfnamefont {J.~M.}\ \bibnamefont {Rubi}},\ and\ \bibinfo {author} {\bibfnamefont {A.}~\bibnamefont {P\'erez-Madrid}},\ }\bibfield  {title} {\bibinfo {title} {Giant acceleration of free diffusion by use of tilted periodic potentials},\ }\href@noop {} {\bibfield  {journal} {\bibinfo  {journal} {Physical review letters}\ }\textbf {\bibinfo {volume} {87}},\ \bibinfo {pages} {010602} (\bibinfo {year} {2001})}\BibitemShut {NoStop}%
\bibitem [{\citenamefont {Shinomoto}\ and\ \citenamefont {Kuramoto}(1986)}]{shinomoto1986phase}%
  \BibitemOpen
  \bibfield  {author} {\bibinfo {author} {\bibfnamefont {S.}~\bibnamefont {Shinomoto}}\ and\ \bibinfo {author} {\bibfnamefont {Y.}~\bibnamefont {Kuramoto}},\ }\bibfield  {title} {\bibinfo {title} {Phase transitions in active rotator systems},\ }\href@noop {} {\bibfield  {journal} {\bibinfo  {journal} {Progress of Theoretical Physics}\ }\textbf {\bibinfo {volume} {75}},\ \bibinfo {pages} {1105} (\bibinfo {year} {1986})}\BibitemShut {NoStop}%
\bibitem [{\citenamefont {Adler}(1946)}]{adler1946study}%
  \BibitemOpen
  \bibfield  {author} {\bibinfo {author} {\bibfnamefont {R.}~\bibnamefont {Adler}},\ }\bibfield  {title} {\bibinfo {title} {A study of locking phenomena in oscillators},\ }\href@noop {} {\bibfield  {journal} {\bibinfo  {journal} {Proceedings of the IRE}\ }\textbf {\bibinfo {volume} {34}},\ \bibinfo {pages} {351} (\bibinfo {year} {1946})}\BibitemShut {NoStop}%
\bibitem [{\citenamefont {Reimann}\ \emph {et~al.}(2002)\citenamefont {Reimann}, \citenamefont {Van~den Broeck}, \citenamefont {Linke}, \citenamefont {H\"anggi}, \citenamefont {Rubi},\ and\ \citenamefont {P\'erez-Madrid}}]{reimann2002diffusion}%
  \BibitemOpen
  \bibfield  {author} {\bibinfo {author} {\bibfnamefont {P.}~\bibnamefont {Reimann}}, \bibinfo {author} {\bibfnamefont {C.}~\bibnamefont {Van~den Broeck}}, \bibinfo {author} {\bibfnamefont {H.}~\bibnamefont {Linke}}, \bibinfo {author} {\bibfnamefont {P.}~\bibnamefont {H\"anggi}}, \bibinfo {author} {\bibfnamefont {J.~M.}\ \bibnamefont {Rubi}},\ and\ \bibinfo {author} {\bibfnamefont {A.}~\bibnamefont {P\'erez-Madrid}},\ }\bibfield  {title} {\bibinfo {title} {Diffusion in tilted periodic potentials: Enhancement, universality, and scaling},\ }\href@noop {} {\bibfield  {journal} {\bibinfo  {journal} {Physical Review E}\ }\textbf {\bibinfo {volume} {65}},\ \bibinfo {pages} {031104} (\bibinfo {year} {2002})}\BibitemShut {NoStop}%
\bibitem [{\citenamefont {Gang}\ \emph {et~al.}(1993)\citenamefont {Gang}, \citenamefont {Ditzinger}, \citenamefont {Ning},\ and\ \citenamefont {Haken}}]{gang1993stochastic}%
  \BibitemOpen
  \bibfield  {author} {\bibinfo {author} {\bibfnamefont {H.}~\bibnamefont {Gang}}, \bibinfo {author} {\bibfnamefont {T.}~\bibnamefont {Ditzinger}}, \bibinfo {author} {\bibfnamefont {C.~Z.}\ \bibnamefont {Ning}},\ and\ \bibinfo {author} {\bibfnamefont {H.}~\bibnamefont {Haken}},\ }\bibfield  {title} {\bibinfo {title} {Stochastic resonance without external periodic force},\ }\href@noop {} {\bibfield  {journal} {\bibinfo  {journal} {Physical review letters}\ }\textbf {\bibinfo {volume} {71}},\ \bibinfo {pages} {807} (\bibinfo {year} {1993})}\BibitemShut {NoStop}%
\bibitem [{\citenamefont {Pisarchik}\ and\ \citenamefont {Hramov}(2023)}]{pisarchik2023coherence}%
  \BibitemOpen
  \bibfield  {author} {\bibinfo {author} {\bibfnamefont {A.~N.}\ \bibnamefont {Pisarchik}}\ and\ \bibinfo {author} {\bibfnamefont {A.~E.}\ \bibnamefont {Hramov}},\ }\bibfield  {title} {\bibinfo {title} {Coherence resonance in neural networks: Theory and experiments},\ }\href@noop {} {\bibfield  {journal} {\bibinfo  {journal} {Physics Reports}\ }\textbf {\bibinfo {volume} {1000}},\ \bibinfo {pages} {1} (\bibinfo {year} {2023})}\BibitemShut {NoStop}%
\bibitem [{\citenamefont {Lindner}\ \emph {et~al.}(2001)\citenamefont {Lindner}, \citenamefont {Kostur},\ and\ \citenamefont {Schimansky-Geier}}]{lindner2001optimal}%
  \BibitemOpen
  \bibfield  {author} {\bibinfo {author} {\bibfnamefont {B.}~\bibnamefont {Lindner}}, \bibinfo {author} {\bibfnamefont {M.}~\bibnamefont {Kostur}},\ and\ \bibinfo {author} {\bibfnamefont {L.}~\bibnamefont {Schimansky-Geier}},\ }\bibfield  {title} {\bibinfo {title} {Optimal diffusive transport in a tilted periodic potential},\ }\href@noop {} {\bibfield  {journal} {\bibinfo  {journal} {Fluctuation and Noise Letters}\ }\textbf {\bibinfo {volume} {1}},\ \bibinfo {pages} {R25} (\bibinfo {year} {2001})}\BibitemShut {NoStop}%
\bibitem [{\citenamefont {Pikovsky}\ \emph {et~al.}(2001)\citenamefont {Pikovsky}, \citenamefont {Rosenblum},\ and\ \citenamefont {Kurths}}]{pikovsky2001synchronization}%
  \BibitemOpen
  \bibfield  {author} {\bibinfo {author} {\bibfnamefont {A.}~\bibnamefont {Pikovsky}}, \bibinfo {author} {\bibfnamefont {M.}~\bibnamefont {Rosenblum}},\ and\ \bibinfo {author} {\bibfnamefont {J.}~\bibnamefont {Kurths}},\ }\bibfield  {title} {\bibinfo {title} {Synchronization},\ }\href@noop {} {\bibfield  {journal} {\bibinfo  {journal} {Cambridge university press}\ }\textbf {\bibinfo {volume} {12}} (\bibinfo {year} {2001})}\BibitemShut {NoStop}%
\bibitem [{\citenamefont {Kuramoto}\ and\ \citenamefont {Kuramoto}(1984)}]{kuramoto1984chemical}%
  \BibitemOpen
  \bibfield  {author} {\bibinfo {author} {\bibfnamefont {Y.}~\bibnamefont {Kuramoto}}\ and\ \bibinfo {author} {\bibfnamefont {Y.}~\bibnamefont {Kuramoto}},\ }\href@noop {} {\emph {\bibinfo {title} {Chemical turbulence}}}\ (\bibinfo  {publisher} {Springer},\ \bibinfo {year} {1984})\BibitemShut {NoStop}%
\bibitem [{\citenamefont {Womelsdorf}\ and\ \citenamefont {Fries}(2007)}]{womelsdorf2007role}%
  \BibitemOpen
  \bibfield  {author} {\bibinfo {author} {\bibfnamefont {T.}~\bibnamefont {Womelsdorf}}\ and\ \bibinfo {author} {\bibfnamefont {P.}~\bibnamefont {Fries}},\ }\bibfield  {title} {\bibinfo {title} {The role of neuronal synchronization in selective attention},\ }\href@noop {} {\bibfield  {journal} {\bibinfo  {journal} {Current opinion in neurobiology}\ }\textbf {\bibinfo {volume} {17}},\ \bibinfo {pages} {154} (\bibinfo {year} {2007})}\BibitemShut {NoStop}%
\bibitem [{\citenamefont {Fell}\ and\ \citenamefont {Axmacher}(2011)}]{fell2011role}%
  \BibitemOpen
  \bibfield  {author} {\bibinfo {author} {\bibfnamefont {J.}~\bibnamefont {Fell}}\ and\ \bibinfo {author} {\bibfnamefont {N.}~\bibnamefont {Axmacher}},\ }\bibfield  {title} {\bibinfo {title} {The role of phase synchronization in memory processes},\ }\href@noop {} {\bibfield  {journal} {\bibinfo  {journal} {Nature reviews neuroscience}\ }\textbf {\bibinfo {volume} {12}},\ \bibinfo {pages} {105} (\bibinfo {year} {2011})}\BibitemShut {NoStop}%
\bibitem [{\citenamefont {Singer}\ \emph {et~al.}(1993)\citenamefont {Singer} \emph {et~al.}}]{singer1993synchronization}%
  \BibitemOpen
  \bibfield  {author} {\bibinfo {author} {\bibfnamefont {W.}~\bibnamefont {Singer}} \emph {et~al.},\ }\bibfield  {title} {\bibinfo {title} {Synchronization of cortical activity and its putative role in information processing and learning},\ }\href@noop {} {\bibfield  {journal} {\bibinfo  {journal} {Annual review of physiology}\ }\textbf {\bibinfo {volume} {55}},\ \bibinfo {pages} {349} (\bibinfo {year} {1993})}\BibitemShut {NoStop}%
\bibitem [{\citenamefont {Remlein}\ \emph {et~al.}(2022)\citenamefont {Remlein}, \citenamefont {Weissmann},\ and\ \citenamefont {Seifert}}]{remlein2022coherence}%
  \BibitemOpen
  \bibfield  {author} {\bibinfo {author} {\bibfnamefont {B.}~\bibnamefont {Remlein}}, \bibinfo {author} {\bibfnamefont {V.}~\bibnamefont {Weissmann}},\ and\ \bibinfo {author} {\bibfnamefont {U.}~\bibnamefont {Seifert}},\ }\bibfield  {title} {\bibinfo {title} {Coherence of oscillations in the weak-noise limit},\ }\href@noop {} {\bibfield  {journal} {\bibinfo  {journal} {Physical Review E}\ }\textbf {\bibinfo {volume} {105}},\ \bibinfo {pages} {064101} (\bibinfo {year} {2022})}\BibitemShut {NoStop}%
\bibitem [{\citenamefont {\ifmmode~\check{S}\else \v{S}\fi{}iler}\ \emph {et~al.}(2018)\citenamefont {\ifmmode~\check{S}\else \v{S}\fi{}iler}, \citenamefont {Ornigotti}, \citenamefont {Brzobohat\'y}, \citenamefont {J\'akl}, \citenamefont {Ryabov}, \citenamefont {Holubec}, \citenamefont {Zem\'anek},\ and\ \citenamefont {Filip}}]{vsiler2018diffusing}%
  \BibitemOpen
  \bibfield  {author} {\bibinfo {author} {\bibfnamefont {M.}~\bibnamefont {\ifmmode~\check{S}\else \v{S}\fi{}iler}}, \bibinfo {author} {\bibfnamefont {L.}~\bibnamefont {Ornigotti}}, \bibinfo {author} {\bibfnamefont {O.}~\bibnamefont {Brzobohat\'y}}, \bibinfo {author} {\bibfnamefont {P.}~\bibnamefont {J\'akl}}, \bibinfo {author} {\bibfnamefont {A.}~\bibnamefont {Ryabov}}, \bibinfo {author} {\bibfnamefont {V.}~\bibnamefont {Holubec}}, \bibinfo {author} {\bibfnamefont {P.}~\bibnamefont {Zem\'anek}},\ and\ \bibinfo {author} {\bibfnamefont {R.}~\bibnamefont {Filip}},\ }\bibfield  {title} {\bibinfo {title} {Diffusing up the hill: Dynamics and equipartition in highly unstable systems},\ }\href@noop {} {\bibfield  {journal} {\bibinfo  {journal} {Physical review letters}\ }\textbf {\bibinfo {volume} {121}},\ \bibinfo {pages} {230601} (\bibinfo {year} {2018})}\BibitemShut {NoStop}%
\bibitem [{\citenamefont {Friston}(2010)}]{friston2010free}%
  \BibitemOpen
  \bibfield  {author} {\bibinfo {author} {\bibfnamefont {K.}~\bibnamefont {Friston}},\ }\bibfield  {title} {\bibinfo {title} {The free-energy principle: a unified brain theory?},\ }\href@noop {} {\bibfield  {journal} {\bibinfo  {journal} {Nature reviews neuroscience}\ }\textbf {\bibinfo {volume} {11}},\ \bibinfo {pages} {127} (\bibinfo {year} {2010})}\BibitemShut {NoStop}%
\bibitem [{\citenamefont {Harris}\ \emph {et~al.}(2012)\citenamefont {Harris}, \citenamefont {Jolivet},\ and\ \citenamefont {Attwell}}]{harris2012synaptic}%
  \BibitemOpen
  \bibfield  {author} {\bibinfo {author} {\bibfnamefont {J.~J.}\ \bibnamefont {Harris}}, \bibinfo {author} {\bibfnamefont {R.}~\bibnamefont {Jolivet}},\ and\ \bibinfo {author} {\bibfnamefont {D.}~\bibnamefont {Attwell}},\ }\bibfield  {title} {\bibinfo {title} {Synaptic energy use and supply},\ }\href@noop {} {\bibfield  {journal} {\bibinfo  {journal} {Neuron}\ }\textbf {\bibinfo {volume} {75}},\ \bibinfo {pages} {762} (\bibinfo {year} {2012})}\BibitemShut {NoStop}%
\end{thebibliography}%

\end{document}